# Grain-size dependent electric-field induced structural changes in relaxor-ferroelectric based unclamped piezoelectric grains

Bhoopesh Mahale, Rishikesh Pandey, Naveen Kumar and Rajeev Ranjan[*]

*Department of Materials Engineering, Indian Institute of Science Bangalore-560012, India.*

## Abstract

Polymer-piezoceramic 0-3 composites combine the flexibility of the polymers and the excellent piezoelectric properties of the ferroelectric based ceramic. While grain size of the ceramic powder is one of the important considerations in the fabrication of such composites, a correlation relating poling field induced structural changes and its possible influence on the overall piezoelectric response of the composite is still lacking. In this paper, we examine this issue on a 0-3 piezo-composite comprising of a ceramic powders of a low-lead piezoelectric alloy $(x)Bi(Ni_{1/2}Zr_{1/2})O_3$-$(1-x)PbTiO_3$ in proximity of its morphotropic phase boundary, and polyvinylidene fluoride (PVDF) as the polymer component. Composites were fabricated by fixing the volume fraction of the ceramic while varying the grain size. We found a non-monotonic variation in the piezo-response as a function of grain size. Structural analysis before and after poling of the piezo-composites revealed evidence of poling induced cubic-like to tetragonal irreversible transformation, the extent of which is dependent on the grain size. Our findings suggest that the non-monotonic grain size dependence of the composite's piezoelectric response is associated with inducing a coexistence of long ranged and short ranged ferroelectric domains in the ferroelectric grains of the composite by the poling field. Our observations contrast a conjecture reported in the past which attributed a similar non-monotonic grain size dependent piezoelectric behavior to the larger grains becoming off-stoichiometric.

*ranjanrajeeb@gmail.com

## I. Introduction

Piezoelectric materials in the form of single crystals, ceramic, polymers, and composites continue to attract attention of the scientific and technological community because of their great technological significance. In general, while advanced piezoelectric ceramics have the advantage of showing large piezoelectric charge coefficient (*d*), dielectric constant, and mechanical coupling coefficient, the usually have disadvantages associated with large acoustic impedance, lower voltage constant (*g*) and high stiffness. This limits there use in flexible sensing applications, hydrophones and some medical devices.[1-3] On the other hand, although piezoelectric polymers such as polyvinylidene fluoride (PVDF) exhibit high piezoelectric voltage constant, low acoustic characteristic impedance and flexibility, they have low piezoelectric charge coefficient, low dielectric constant and mechanical coupling coefficient.[1-3] For over three decades now, attempts have been made to overcome the limitations of the ceramic and polymer piezoelectrics by developing polymer-ceramic composite piezoelectric materials.[1-5] Composite materials exhibit better performance over pure ceramics by incorporating good elastic properties and durability. 0-3 connectivity (Ceramic filler surrounded by polymer matrix) introduced by Newnham *et al.*[5] is the most popular among researchers because of its ease of fabrication, ease of variation in properties by changing the filler volume fraction in wide range, and for their flexibility to adapt in different shapes.[6-8] Several groups have studied the 0-3 composites extensively using various polymer types such as epoxy resin, liquid crystalline thermosets (LCT)[7-9] and thermoplastic polymers like PVDF.[3,4,10-15] In majority of the studies, the most popular among the piezoelectrics ceramics lead zirconate titanate $Pb(Zr_{1-x} Ti_x)O_3$ (PZT) has been used as the piezoceramic component.[7-13] There has been some reports on Barium Titanate ($BaTiO_3$)-polymer nanocomposites as well.[14-16]

Fabrication of a good polymer-ceramic piezocomposite requires homogeneous distribution of the ceramic particles in the polymer. This is relatively easy to attain when ceramic grains are small. However, the very piezoelectric property is compromised by grain size reduction.[17-20] In view of this, the best piezoelectric property of a polymer-piezoceramic composite can be expected to occur for an optimum size of the ceramic grains. Most often the grain size variation of the piezoceramic powder is carried out by varying the calcination/sintering temperature. Except for BaTiO$_3$-based piezoceramics, many of the other interesting piezoelectric systems such as Pb(Zr, Ti)O$_3$, (K, Na)NbO$_3$, Na$_{1/2}$Bi$_{1/2}$TiO$_3$ have volatile elements such as Na, K, Bi, Pb, which increases the possibility of unintentional change in the stoichiometry of the piezoceramic particles when they are heated at high temperatures for increasing their grain size. Some groups have shown that the piezoelectric response of the composite decreases above certain grain size, and have speculated this to arise from such an unintentional alteration in the stoichiometry.[7] While this speculation appears plausible, its validity has not been examined. Further, there is lack of alternative viewpoints to understand the size dependent non-monotonic piezoelectric response in 0-3 piezo-composite. In this paper, we have approached the issue from a structure-property correlation standpoint. A similar approach was attempted before for a BaTiO$_3$-PVDF piezo-composite,[16] but there is no clarity on structural changes accompanying the non-monotonic size dependent piezo-response. Further progress in this field would require not only understanding of the factors (including grain size and their morphology) which optimizes the processing conditions for best piezoelectric response, but also a detailed understanding of the structural changes accompanying the grain sizes in such composites, and most importantly after the specimen has been poled. This aspect is important in view of some of the recent reports which suggests phase electric field induced structural transformations in the high performance piezoceramics.[21-26] In this paper, we investigate this aspect in detail and attempt to correlate the property of the composite with the nature of the electric field induced structural transformations, if any. For our study, we used (x)Bi(Ni$_{1/2}$Zr$_{1/2}$)O$_3$-(1-x)PbTiO$_3$ (BNZ-PT) as the piezoceramic material and PVDF as the polymer matrix. BNZ-PT exhibit piezoelectric charge constant (d$_{33}$) ~ 400pC/N,[27] a value comparable to Pb(Zr$_x$Ti$_{1-x}$)O$_3$(PZT) in the vicinity of its MPB. This material has nearly half the lead (Pb) content compared to the conventional piezoelectric PZT, and is therefore relatively less toxic. Further, analogous to PZT, BNZ-PT has Bi and Pb as volatile elements and therefore provides an opportunity to compare our results with those reported for PZT-polymer piezo-composite, and examine if off-stoichiometry could be a determining factor in influencing the piezoelectric response of such composites. In our study, we chose a morphotropic phase composition of BNZ-PT and varied the grain size of the ceramic while keeping its volume fraction constant while fabricating the composite with the polymer (PVDF). While, in conformity with some of the reports in the past, our composites also show a non-monotonic piezoelectric response as a function of average grain size, we found that the non-monotonic piezoelectric response is primarily associated with the extent of electric field induced irreversible cubic-like to tetragonal transformation after poling the composite, and not due to setting in of off-stoichiometry, if any in the large sized grains.

## II. Experimental

### A. Piezo Ceramic Powder Preparation

The morphotropic phase boundary (MPB) of (x)BNZ-(1-x)PT is reported to occur at x ~ 0.41.[27] We prepared this composition by conventional solid state synthesis technique. Raw chemicals (Reagent grade) in the form of oxides viz. $Bi_2O_3$ (99%), NiO (99%), $ZrO_2$ (99.9%), $TiO_2$ (99.8%), and PbO (99.9%) (All from Alfa Aesar) were dried in hot air oven at 150°C for 2h. All the chemicals weighed and mixed according to stoichiometric formulas of each composition. The mixtures were then ball milled in agate vials in acetone medium at 200 rpm for 8 h. Ball milled powder was dried and calcined in alumina crucibles at different temperatures. The calcined powders were then ball milled in ethanol medium to break the agglomeration.

### B. Composite fabrication

We, first, optimized the ceramic content in the composite, by fabricating large number of films with different loading of the ceramic in the polymer. A homogenous distribution of the ceramic powder was obtained for 50 - 60 vol % of the ceramic content. Above this fraction, the composite with submicron grain powders tend to be brittle in nature. On the other hand, below this limit the composite with large sized grains showed non-homogenous distribution. Accordingly, for sake of a meaningful comparative study as a function of grain size, we chose 55 vol % ceramic content as optimum. This ratio ensured a combination of good flexibility and homogeneous distribution of the ceramic powders for all grain sizes. The PVDF solution was prepared by dissolving 20 wt% of PVDF (Kynar-761 with Mw = 440 000 g /mol, obtained from Arkema Inc.) in N,N-Dimethyleformamide (DMF) solvent. Mixture was constantly stirred on magnetic stirrer at 50°C for 30min to dissolve the PVDF. Predetermined amount of BNZ-PT powder was then added to the PVDF-DMF solution and the slurry was thoroughly stirred. The slurry was poured on a glass substrate and dried in hot air oven at 100°C for 2 h to evaporate the solvent. The composite films thus obtained were placed between hotplates at 170°C and pressed with the pressure around 800psi.  The thickness of the films was found to be in the range 250-300μm. The composites with different grain size of piezoceramic were prepared by keeping the volume fraction of the ceramic powder and the polymer constant. The composite films were electroded with silver paste on both sides. Poling of the films was carried out at 80kV/cm for 1h at room temperature in silicon oil.

## C. Characterization

X-ray diffraction (XRD) patterns of all powder samples were obtained by X-ray diffractometer (Rigaku, Smart Lab) using a monochromatic CuKα (λ=1.5406 Å) radiation. Microstructural characteristics of calcined powder and composites were performed by Scanning Electron Microscopy (SEM) (Quanta). Grain size was measured and analyzed using image J analysis software. However, for the powders calcined at 850 $^o$C and 900 °C, agglomeration of the small grains prevented their distinct identification in the SEM images. We, therefore, estimated the average crystallite sizes of these samples using Scherrer equation. For powder samples calcined at higher temperatures (> 900 $^o$C), the widths of the Bragg peaks were comparable to the instrumental width of diffractometer, and hence their crystallite size were not estimated using the Scherrer equation. Piezoelectric charge coefficient of the poled composites was measured using a piezometer (Piezotest PM-300).

## III. Results
## A. Grain size dependent piezoresponse of the composite

Figure 1 shows the x ray powder diffraction (XRD) patterns of BNZ-PT powders calcined at different temperatures. All the samples exhibit formation of perovskite phase with a cubic-like structure as there was no splitting in the Bragg peaks. This is consistent with a similar observation reported before for this alloy system.[27] The significant broadening of profiles at low calcination temperature suggests small crystallite size. From the SEM images of BNZ-PT powders calcined at different temperatures, shown in Fig. 2, it is evident that the grain size increases with increasing the calcination temperature. For powders calcined at 850 $^o$C and 900 $^o$C, the individual grains could not be resolved in the SEM images. Their average crystallite sizes were estimated using the Scherrer equation. For correct estimation of crystallite size, we collected powder diffraction data at 400 $^o$C, i.e. in the cubic paraelectric state of the specimen, to get rid of the possible broadening in the diffraction peaks due to small lattice strain associated with the ferroelectric state at room temperature. For the powders calcined at higher temperatures (> 900 C), the boundaries of the grains could be seen and image J software was used to get the grain size distribution. The average particle size of the powder calcined at different temperatures is given in Table I.

Figure 3 shows the cross-sectional SEM images of the polymer-piezoceramic composites. On this scale, the piezoceramic grains in all the composites appear to be uniformly distributed across the volume of the specimen. Figure 4 shows the variation of longitudinal piezoelectric charge coefficient ($d_{33}$) of the composite films as a function of average grain size of the ceramic powder. A non-monotonic variation of $d_{33}$ with size, with a maximum at ~ 1.5 µm is evident from this plot. We may note that for the same poling field (80 kV/cm), the PVDF did not give a noticeable $d_{33}$, suggesting that the piezoresponse of the composite is almost entirely contributed by the ceramic grains. The trend observed by us is similar to what was reported earlier by Babu *et al.* for a 0-3 PZT-polymer piezo-composite.[7] Babu *et al.* attributed the initial increase in the piezoelectric response to increasing grain size, and the subsequent decrease to lead loss in their PZT ceramic grains. Similar to PZT, our specimens too have volatile elements such as Pb and Bi which can evaporate when heating the specimen at high temperatures. We synthesized bulk ceramic pellets of BNZ-PT by sintering at temperatures 1000 °C, 1050 °C and 1120°C, and found the $d_{33}$ of the pellet sintered at 1120°C to be higher ($d_{33}$ = 335 pC/N) than those sintered at lower temperatures ($d_{33}$ = 314 pC/N sintered at 1050°C; $d_{33}$ = 289 pC/N sintered at 1000°C). This confirms that the large sized BZN-PT piezoceramic exhibit larger piezoelectric response, and that the increased volatility of the elements, if any, has not compromised the piezoelectric response of ceramics with the largest grain size.

**B. Poling induced irreversible structural change**

Since experimentally the piezoelectric response is measured after subjecting the specimen to poling field, it is also important to analyze the structural state of the piezoelectric particles after poling. In this context, it may be noted that the relaxor ferroelectric systems NBT and NBT-BT exhibit considerable change in the average global structure after poling.[21-24] Figure 5 shows the {200}$_{pc}$ pseudocubic XRD Bragg profiles of BNZ-PT-PVDF composite films before and after poling at 80kV/cm. A dramatic change in the shape of the Bragg profiles after poling suggests a field induced irreversible change in the global structure. The splitting of the {200}$_{pc}$ into two peaks in the XRD pattern of the poled composite comprising of large grains ( 2.2 and 3.4 µm) suggests a poling induced cubic-like to tetragonal transformation. In an ideal powder diffraction pattern of a tetragonal perovskite ferroelectric, the intensity of the (002) tetragonal Bragg peak is smaller (less than half) than that of the (200) tetragonal Bragg peak. The higher intensity of the (002) peak than the (200) peak in the pattern of the powder synthesized at 1120°C (grain size 3.4 µm) is due to the poling induced reorientation of the tetragonal domains. The extent of tetragonal domain reorientation is noticeably reduced in the grains of the piezoceramic synthesized at lower temperature 1050°C (grain size 2.2 µm), as evident from the nearly equal intensities of (002) and (200) peaks. The poled grains of the ceramic synthesized at 1000°C (grain size 1.5µm) show coexistence of the cubic-like and the tetragonal peaks. As evident from Fig. 4, this composite exhibits the maximum piezo response among others. The Bragg peaks are considerably broadened for piezoceramics synthesized below 1000°C. These powder shows grain sizes in the submicron range. The piezo charge coefficient (32 pC/N at 55 vol % of ceramic) in the present work is comparable to the ones reported in the literature e.g. 42 pC/N (50vol% PZT in liquid crystal thermoset/Poly amide),[7] ~17 pC/N (40vol % PZT in LCT),[8] 48 pC/N (67 vol% PZT in PVDF),[10] 14 pC/N (50 vol% PZT in PVDF),[12] ~22 pC/N (60 vol% PZT in PVDF),[13]

We may mention that a new series of composite samples were also made and the non-monotonic trend in the piezoelectric response was found to be reproducible, lending authenticity to the results. We discuss the implications of these results in the next section.

## IV. Discussion

The partial transformation of the cubic-like phase to tetragonal after poling creates a two-phase scenario which appears to be equivalent to the composition induced two-phase state in morphotropic phase boundary (MPB) compositions of ferroelectric alloy systems such as in PZT.[28] In the well-known piezoelectric alloy PZT, the MPB is characterized by coexistence of rhombohedral (R3m) and tetragonal (P4mm) ferroelectric phases.[28] In contrast, two-phase state in BNZ-PT is cubic-like + tetragonal (P4mm). The field driven cubic-like to MPB transformation observed in our system is analogous to what has reported earlier in the critical composition (x = 0.06) of the relaxor ferroelectric system $(1-x)Na_{1/2}Bi_{1/2}TiO_3 - (x)BaTiO_3$(NBT-6BT).[23,24] Garg *et al.* have shown that the average cubic-like phase of transforms to tetragonal and rhombohedral phases after poling.[23] The onset of the rhombohedral phase after poling can be expected in NBT-6BT as the parent compound NBT shows a monoclinic-like to rhombohedral transformation on poling.[21,22] On the contrary, BNZ does not crystallize into a perovskite phase under ambient pressure conditions. The cubic-like phase in BNZ-PT appears to be a manifestation of significant effect of local positional disorder on the global average structure of the system. In BNZ-PT, Bi is expected to occupy the Pb sites and Zr and Ni the Ti-sites of $PbTiO_3$. Bi is known to enhance ferroelectric distortion when part of a perovskite structure as in $BiFeO_3$[29] and $BiFeO_3$-$PbTiO_3$.[30] However, the ferroelectric inactive ions Zr and Ni will tend to suppress ferroelectricity and lead to miniaturization of the ferroelectric domains and relaxor ferroelectric behaviour.[31] Diffraction techniques such as x-ray and neutron powder diffraction give only the average structural information over a large (global) length scale. If the system is structurally homogeneous, the structures on the global and local (on the scale of unit cell) would be same. If not, the average structure becomes dependent on the length scale of the probing technique. Using pair distribution function analysis, Usher *et al.* have recently shown that the cubic-like average structure of (1-x)$BaTiO_3$-(x)$Bi(Zn_{1/2}Nb_{1/2})O_3$ appear as tetragonal when examined on a smaller length scale.[32] In such a scenario, what is perceived as a field induced cubic-like to tetragonal phase transformation on the global scale (XRD) in our piezoceramic grains can be interpreted as increase in the coherence length of tetragonal domains by the poling field. The cubic-like + tetragonal two-phase state in our case therefore represents a coexistence of long and short coherence lengths of tetragonal domains. This scenario may as well be interpreted as a coexistence of relaxor and normal ferroelectric regions inside the grains. The large piezoelectric response in the seemingly two-phase state appears to be associated with this kind of polar-structural heterogeneity induced by the poling field in the optimized sized grains. In the scenario, wherein the poling field transforms completely the cubic-like state to tetragonal, as in the large sized grains, the degree of the polar-structural heterogeneity is reduced. Our explanation is qualitatively different from the viewpoint which attributes a similar decrease in the piezoresponse of PZT-polymer composite to unintentional off-stoichiometry in the large grains because of increased volatility of Pb at high temperatures. As stated above, like PZT, our large grain piezoceramic synthesized at 1120°C might also have become off-stoichiometric to some extent due to volatilization of Pb and Bi, however the bulk ceramic shows higher piezoresponse than those synthesized at lower temperatures, ruling out off-stoichiometry, if any, as the primary factor for the decrease in the piezoelectric response.

The decrease in piezoelectric response below the optimum grain size can be attributed to the factors which tend to suppress ferroelectricity in small size ferroelectrics. It is well known that the average structure of $BaTiO_3$ transforms from tetragonal (*P4mm*) to cubic below 0.1μm.[17-19] $PbTiO_3$ also shows a similar phenomenon below ~15 nm.[20] There are different explanations for the suppression of ferroelectricity in small sized ferroelectrics. Some groups have attempted to explain this phenomenon by invoking the concept of extrapolation length to characterize the different nature of polarization at the particle surface than that of the interior.[33] The role of depolarizing field has also been considered.[34] In some studies surface tension has been argued to be the destabilizing factor.[35,36] Geneste *et. al.* have considered bond contraction in the few atomic layers of the surface as a possible reason to suppress ferroelectricity in small size ferroelectric particles.[37] Presence of dead surface layers and their enhanced volume fraction with decreasing size is also considered as a contributing factor.[38] While we cannot speculate on dominant mechanism among these possibilities to explain the decreases in the piezoelectric response below the optimal grain size in our case, a careful examination of the XRD patterns of the poled composites with submicron grains seems to offers a plausible correlation. Although similar to the piezoceramic with optimum grain size, the submicron grains of BNZ-PT also show evidence of field induced cubic-like to tetragonal transformation, the $\{200\}_{pc}$ profiles of the sub-optimal sized grains are considerably broadened as compared to that of the optimal sized grains. This suggests that field induced tetragonal domains are so not well developed in the sub-optimal sized grains when compared with those with optimal sized grains and above. The submicron grains are therefore not as efficiently poled as the grains with optimal size and above. Above the optimal size, although poling becomes increasingly efficient with increasing grain size, the grains loose the desired structural and polar heterogeneity for enhanced piezoelectric response. The enhanced piezo-response in the optimal sized grains is therefore a result of fulfillment of conditions wherein poling filed irreversibly stabilizes the coexistence of  long range (ferroelectric) and short ranged (relaxor ferroelectric) tetragonal domains.

**V. Conclusions**

In summary, we investigated the piezoelectric response of a 0-3 piezoelectric –polymer composite as a function of grain size of the piezoceramic grains. We found that composites fabricated by mixing a fixed volume fraction of the MPB composition of BZN-PT piezoceramic powders of varying grain sizes with PVDF show a non-monotonic grain size dependent piezoelectric response ($d_{33}$). The maximum response was found for the optimal average grain size $\sim 1.5\mu m$. We investigated the structural states of the ceramic powders in the composites before and after electric poling, and found evidence of poling induced cubic-like to tetragonal transformation. This transformation is nearly complete in grains above the optimal size, and partial in the optimal sized grains and below. Our results suggest that the relatively high piezoelectric response of the composite with optimal size grains is associated with the poling field induced structural heterogeneity wherein tetragonal domains of long coherence length (ferroelectric phase) coexists with a minor fraction short coherence length. While this scenario seems to persist even below the optimal size (submicron grains), they show decrease in the piezoelectric response due to the limitations imposed by the small size on the ability to develop long ranged tetragonal regions on application of the poling field.

**ACKNOWLEDGEMENTS**

R. Ranjan gratefully acknowledges the Science and Engineering Research Board (SERB) of the Ministry of Science and Technology, Govt. of India, for financial support (Grant No. EMR/2016/001457). R. Pandey gratefully acknowledges SERB for the award of National Post-doctoral Fellowship.

TABLE I. Variation of average grain size with calcination temperature.

| Calcination Temperature (°C) | Average grain size (μm) |
|---|---|
| 850 | 0.21 (by Scherrer equation) |
| 900 | 0.39 (by Scherrer equation) |
| 950 | 0.81 (by SEM) |
| 1000 | 1.45 (by SEM) |
| 1050 | 2.15 (by SEM) |
| 1120 | 3.36 (by SEM) |

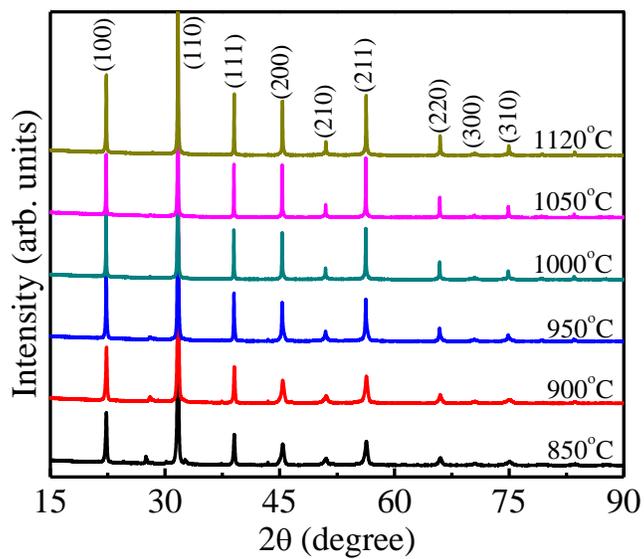

FIG.1. X-ray powder diffraction patterns of powders of BNZ-PT heated at different temperatures. The Miller indices with respect to the pseudo-cubic cell are mentioned above the peaks.

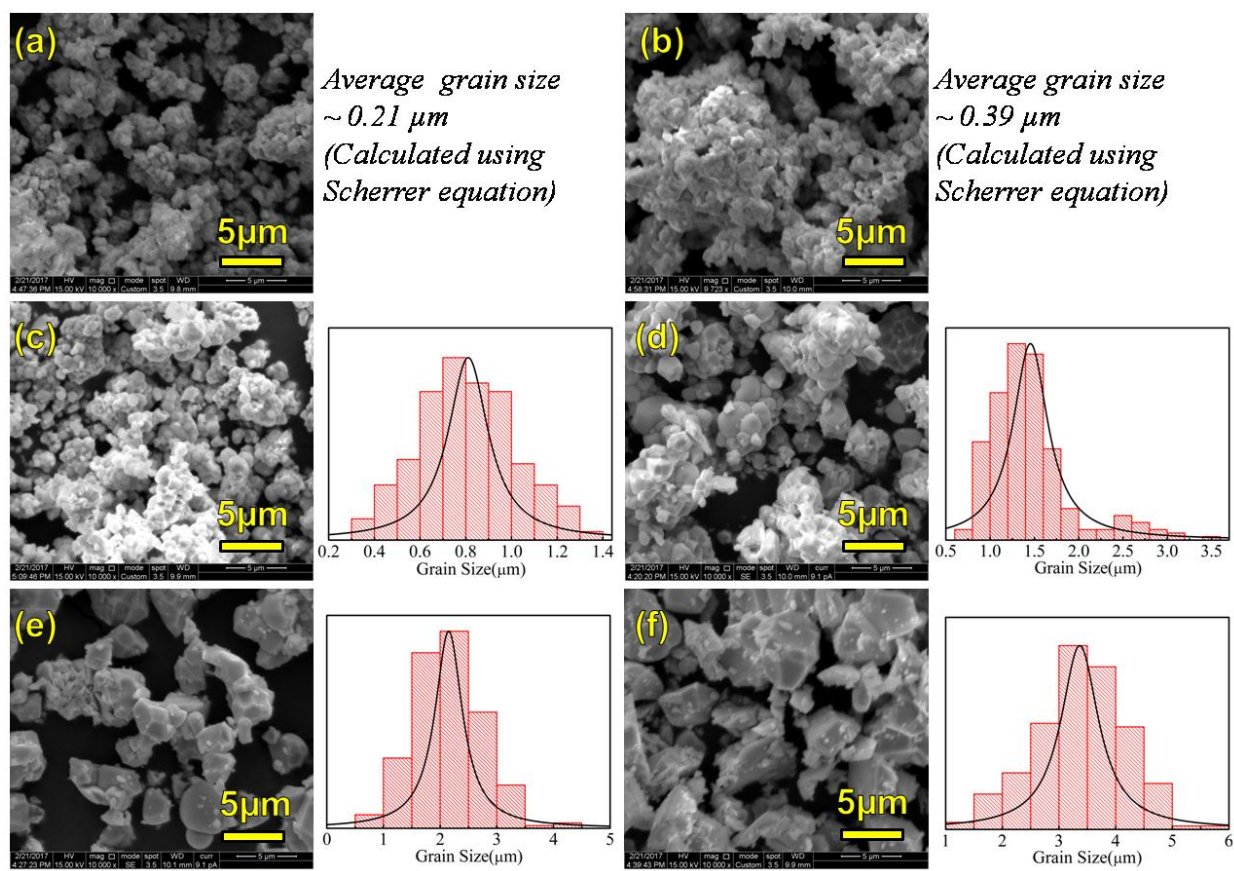

FIG. 2. SEM images of the ceramic powder calcined at (a) 850°C, (b) 900°C, (c) 950°C, (d) 1000°C, (e) 1050°C and (f) 1120°C.

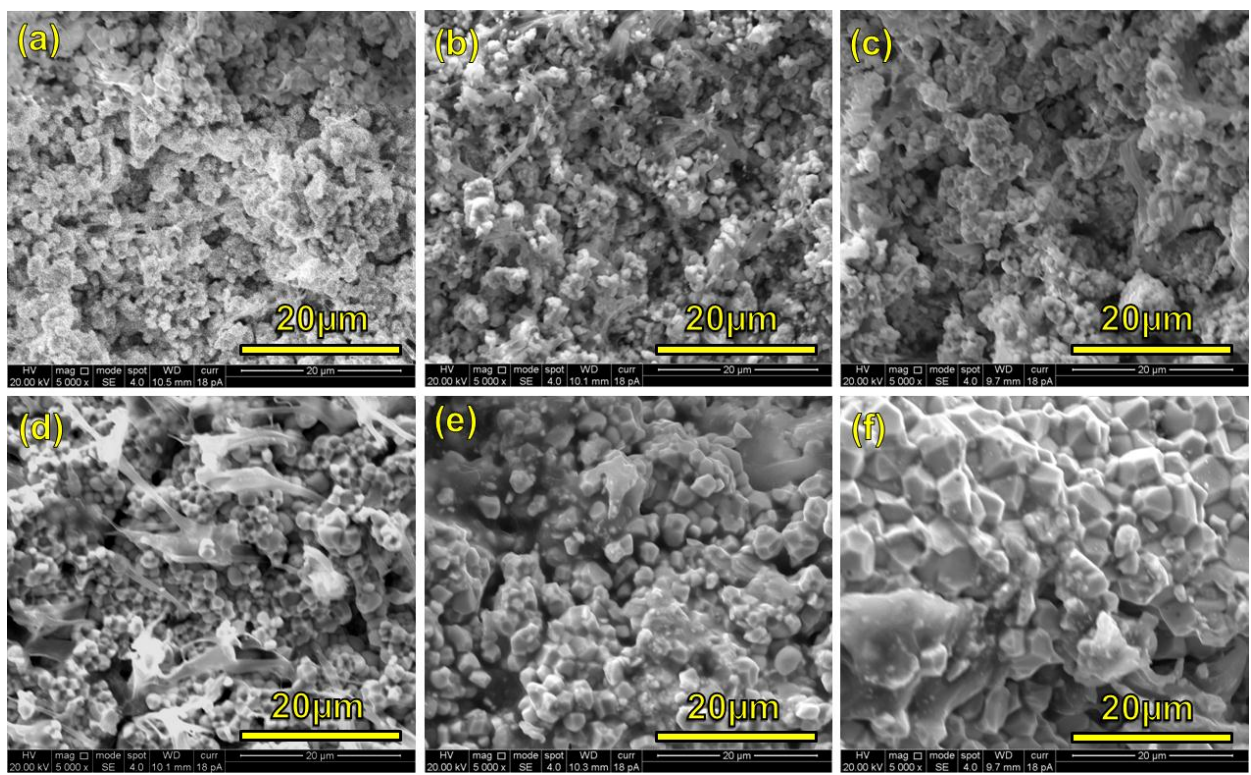

FIG.3. SEM images of cross section of the composite films prepared using ceramic powder calcined at (a) 850°C, (b) 900°C (c) 950°C, (d) 1000°C, (e) 1050°C and (f) 1120°C.

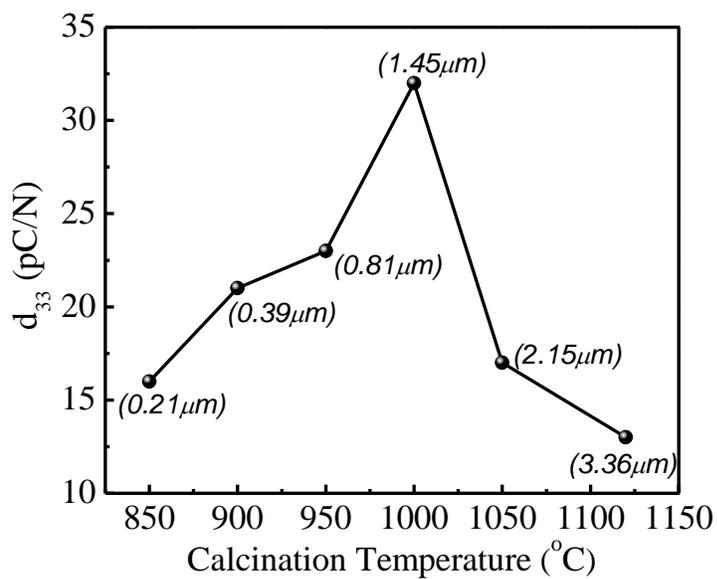

FIG. 4. Variation of piezoelectric charge coefficient ($d_{33}$) of the composite films as a function of calcination temperature of the ceramic powder. The average grain size is also mentioned for each calcination temperature.

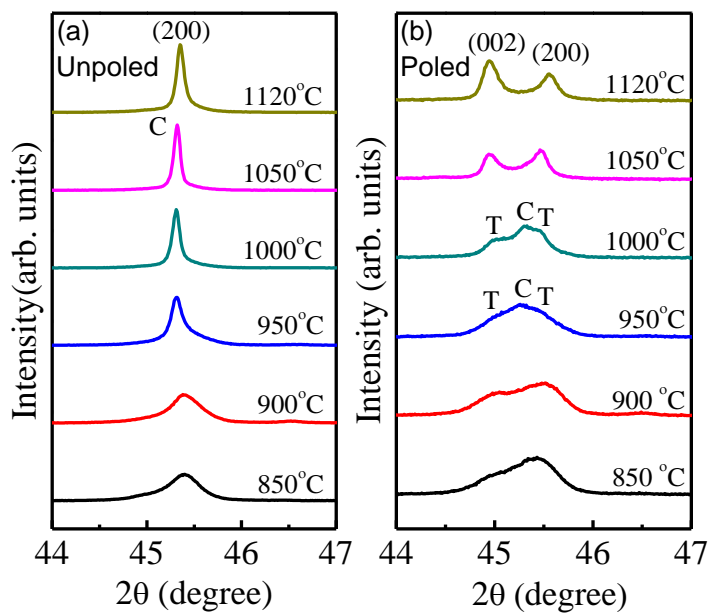

FIG.5. XRD profiles of pseudo-cubic $\{200\}_{pc}$ reflections in the (a) unpoled and (b) poled states of the composite films.